\documentclass[aps,prl,twocolumn,superscriptaddress]{revtex4}
\usepackage{graphicx}
\usepackage{braket}
\usepackage{amsmath}
\usepackage{amssymb}
\usepackage{mathrsfs}
\usepackage{color}
\newcommand{\I}{\mathrm{i}}

\begin{document}


\title{Signatures of electronic structure in bi-circular high-harmonic spectroscopy}


\author{Denitsa Baykusheva}
\thanks{D.B. and S.B. contributed equally to this work.}
\affiliation{Laboratorium f\"ur Physikalische Chemie, ETH Z\"urich,\\
Vladimir-Prelog-Weg 2, 8093 Z\"urich, Switzerland}
\author{Simon Brennecke}
\thanks{D.B. and S.B. contributed equally to this work.}
\affiliation{Laboratorium f\"ur Physikalische Chemie, ETH Z\"urich,\\
Vladimir-Prelog-Weg 2, 8093 Z\"urich, Switzerland}
\author{Manfred Lein}
\affiliation{Institut f\"ur Theoretische Physik, Universit\"at Hannover, 30167 Hannover, Germany}
\author{Hans Jakob W\"orner}
\email[]{hwoerner@ethz.ch}
\homepage[]{www.atto.ethz.ch}
\affiliation{Laboratorium f\"ur Physikalische Chemie, ETH Z\"urich,\\
Vladimir-Prelog-Weg 2, 8093 Z\"urich, Switzerland}

\date{\today}

\begin{abstract}

High-harmonic spectroscopy driven by circularly-polarized laser pulses and their counter-rotating second harmonic is a new branch of attosecond science which currently lacks quantitative interpretations. We extend this technique to the mid-infrared regime and record detailed high-harmonic spectra of several rare-gas atoms. These results are compared with the solution of the Schr\"odinger equation in three dimensions and calculations based on the strong-field approximation that incorporate accurate scattering-wave recombination matrix elements. A quantum-orbit analysis of these results provides a transparent interpretation of the measured intensity ratios of symmetry-allowed neighboring harmonics in terms of (i) a set of propensity rules related to the angular momentum of the atomic orbitals, (ii) atom-specific matrix elements related to their electronic structure and (iii) the interference of the emissions associated with electrons in orbitals co- or counter-rotating with the laser fields. These results provide the foundation for a quantitative understanding of bi-circular high-harmonic spectroscopy.
\end{abstract}


\maketitle

High-harmonic spectroscopy driven by circularly-polarized laser fields superimposed with their counter-rotating second harmonic is a new technique that attracts considerable attention because of its wide application potential, such as the characterization of dynamical symmetries in atoms and molecules \cite{baykusheva16a}. This new research direction has been opened by the pioneering demonstration of high-harmonic generation in such laser fields \cite{eichmann95a} and its theoretical interpretation \cite{long95a,zuo95a,milosevic00a}. The potential of this early work has only recently been fully exploited for the generation of circularly-polarized high-harmonic radiation \cite{fleischer14a}, including its extension to high photon energies \cite{fan15a} and its applications \cite{kfir15a}.
Bi-circular high-harmonic spectroscopy (BHHS) has also received considerable theoretical attention, in particular due to its sensitivity to atomic and molecular symmetry \cite{medisauskas15a,mauger2016,reich16a,hasovic16a,odzak16a,bandrauk16a}, molecular chirality \cite{smirnova15a} and spin polarization \cite{milosevic16a,ayuso16a}. However, both experimental and theoretical results have remained very scarce, such that the sensitivity of BHHS to the various aspects of electronic structure remains largely unknown and it is not clear which theories reach quantitative accuracy.

In this letter, we address these challenges by reporting a joint experimental and theoretical analysis of the fundamental working principles of BHHS in rare-gas atoms. Studying the intensity ratios of neighboring harmonic orders $3q+1$ and $3q+2$ ($q\in\mathbb{N}$) allowed by symmetry, we observe striking differences between the spectra of neon and argon. The experimental results are qualitatively well reproduced by solving the time-dependent Schr\"odinger equation (TDSE) in three dimensions. We generalize our results by developing a model based on the quantum-orbit analysis within the strong-field approximation (SFA) that allows for complex-valued electron trajectories, similar to Ref. \cite{milosevic00a}, but incorporates accurate scattering-wave recombination matrix elements \cite{woerner09a, le09a}. These results are also in good agreement with the experimental data, although significant deviations are found close to the ionization threshold. The analysis of the SFA results allows us to separate the contributions of strong-field ionization and photorecombination to the observed ratios. Our work establishes propensity rules for BHHS based on the angular momentum of atomic orbitals. It additionally reveals the manifestations of orbital-specific radial structures in BHHS. In particular, the sign change of the radial 3p$\rightarrow$d photoionization matrix element of argon with energy, responsible for the Cooper minimum \cite{cooper62a}, is shown to cause a reversal of the relative intensities of neighboring harmonics.

The experimental setup is similar to the one described in Ref. \cite{baykusheva16a}, but is augmented by the capability of generating bi-circular laser fields at wavelengths longer than the standard 800/400-nm of titanium-sapphire lasers. This new development compensates the lower cut-off energies achieved in bi-circular laser fields by an increase of the ponderomotive energy, such that the spectra recorded in argon extend beyond the region of the Cooper minimum. This is achieved by pumping a high-energy optical parametric amplifier (HE-TOPAS, Light Conversion) with up to 6.5 mJ, 30 fs laser pulses centered at 800 nm at a repetition rate of 1 kHz to generate 1.5 mJ pulses with $\sim$ 40~fs duration centered in the vicinity of 1400~nm, that are subsequently frequency doubled to $\sim$700~nm in a nonlinear crystal. We use a Mach-Zehnder interferometer equipped with dedicated dichroic mirrors for each wavelength pair. High-harmonic spectra are generated in a thin supersonic beam generated by expansion of neon or argon through a pulsed nozzle with a diameter of 250 $\mu$m at a stagnation pressure of $\sim$5~bar. The high-harmonic spectra are recorded with a flat-field spectrometer consisting of a concave 1200 lines/mm grating, a microchannel-plate-phosphor-screen assembly and a charge-coupled device camera.

Figure 1 shows a typical bi-circular high-harmonic spectrum recorded in neon using 800/400-nm laser pulses and a spectrum recorded in argon using 1404/702-nm pulses. The different laser parameters, given in the caption of Fig. 1, were chosen to keep the Keldysh parameter (defined on the basis of the fundamental field) constant. Whereas the neon spectrum displays a monotonically decreasing intensity envelope from the ionization potential (dotted line) to the cut-off, the argon spectrum reveals a suppression around photon energies of $\sim$~45~eV, i.e. in the region of the Cooper minimum. We however note already, that this position is lower than the 53-eV position observed in the case of linear HHS \cite{woerner09a}.

\begin{figure}[t]
\includegraphics[width=0.5\textwidth]{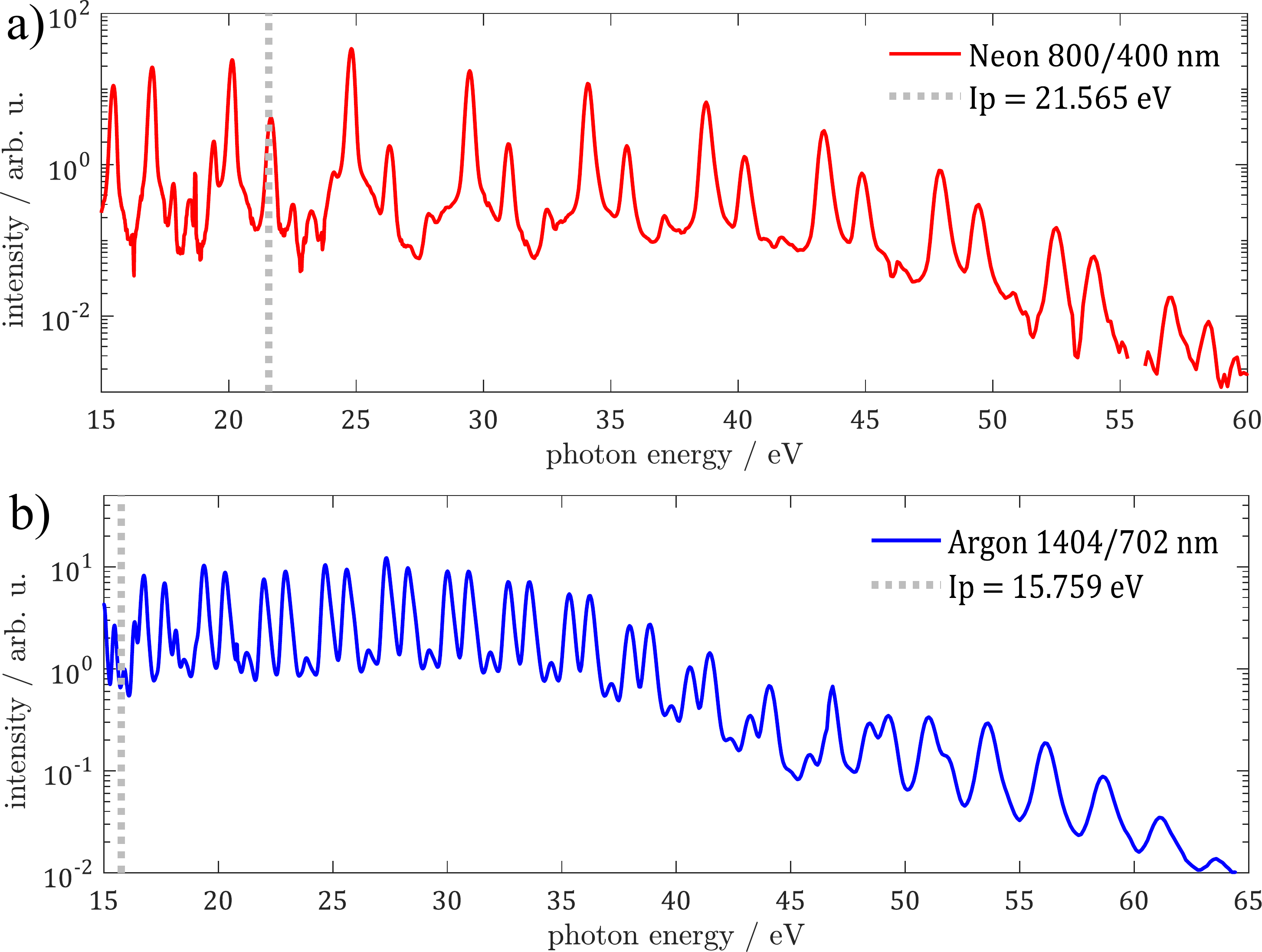}
\caption{(color online) Experimental high-harmonic spectra generated with counter-rotating circularly polarized femtosecond laser pulses in rare-gas atoms. a) High-harmonic spectrum of neon obtained with 800/400~nm pulses with intensities 3.0/1.8$\times 10^{14}$~W/cm$^2$, b) High-harmonic spectrum of argon obtained with 1404/702~nm pulses with peak intensities of 7.2/6.0$\times 10^{13}$~W/cm$^2$. }
\end{figure}

Importantly, BHHS offers an additional observable compared to linear HHS, i.e. the intensity ratios of the neighboring allowed harmonic orders $I(3q+1)/I(3q+2)$. This ratio is a robust observable because it is insensitive to the slow variation of the grating and detector sensitivities with photon energy. It is shown in Fig. 2 as a function of $q$ and the photon energy (using the relation $E=\hbar\omega(3q+1)$ with $\omega$ the fundamental angular frequency). The experimental data are shown as squares with error bars representing twice the standard deviation of multiple measurements taken under nominally identical experimental conditions. Neon displays a very large intensity ratio ($>18$) close to the ionization threshold and a rapid, monotonic decrease of this ratio with increasing photon energy. Argon, in contrast, shows a much richer variation of the ratio with photon energy, with a weak decrease of the ratio from $\sim$29~eV ($q=10$) upwards, followed by an inversion of the ratio at $\sim$40~eV ($q=14$) and a second inversion at $\sim$51~eV ($q=18$). The finding that the ratio is mostly larger than one can be interpreted as due to the less likely absorption of photons from the second harmonic field when its intensity is smaller than the fundamental intensity \cite{dorney17a}. A quantitative understanding must, however, consider the specific atomic structure as explained in the following.

\begin{figure}[t]
\includegraphics[width=0.5\textwidth]{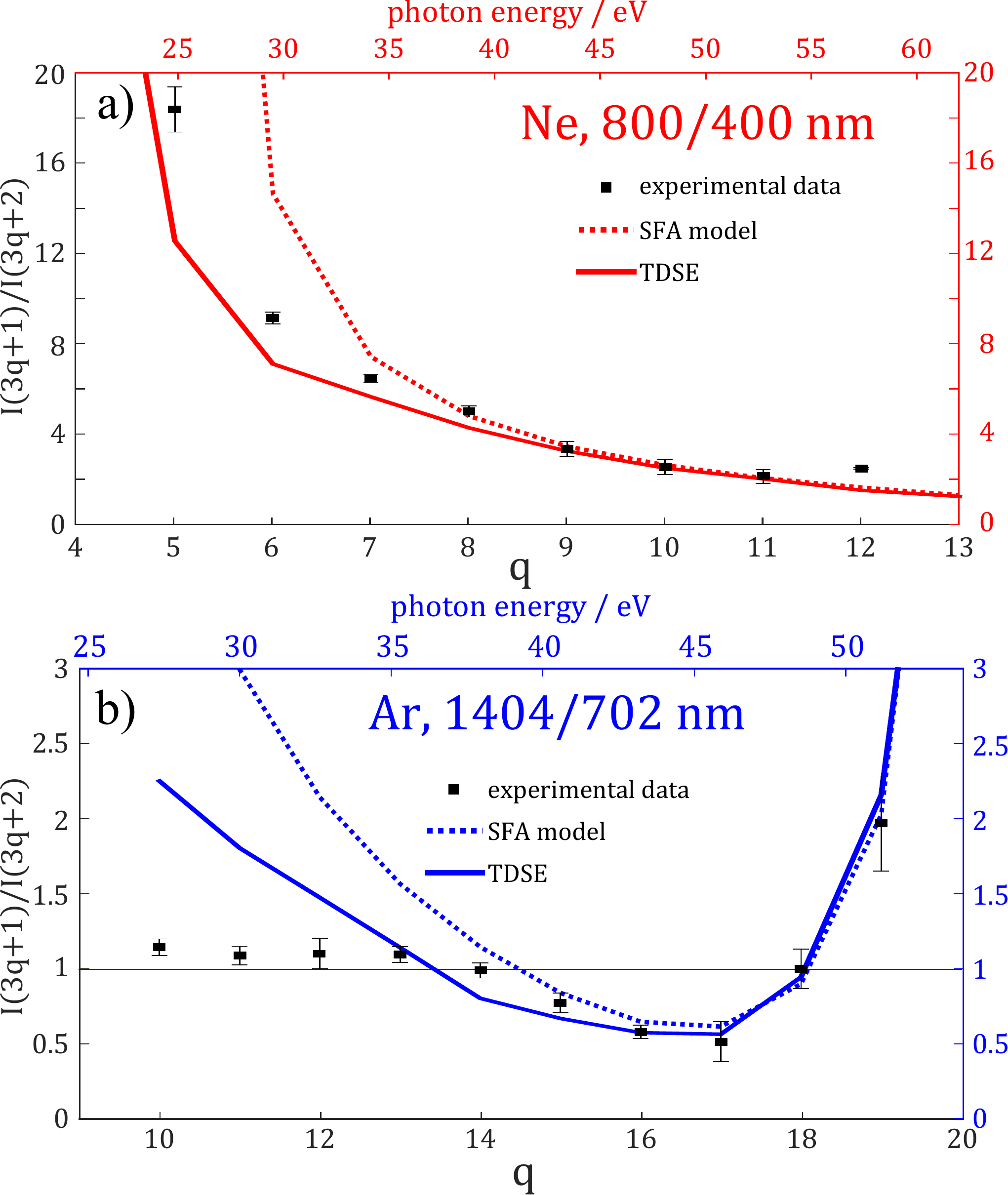}
\caption{(color online) Ratios of integrated intensities of neighboring high-harmonic orders as indicated on the vertical axis, as a function of the integer $q$ for Ne, 800/400~nm (a) and Ar, 1404/702~nm (b). The symbols represent the experimental data. The lines correspond to different theoretical models discussed in the text using the intensities given in the caption of Fig. 1. The theoretical results have been shifted by $+6.2$~eV in the case of Ar (see text for discussion).}
\end{figure}

We now discuss the different theoretical models to which the experimental data will be compared. The TDSE is solved numerically in three dimensions in the length gauge and the single-active-electron approximation. The driving field consists of a circularly polarized fundamental that rotates counterclockwise and a circularly polarized second harmonic that rotates clockwise in the x-y-plane. We use a $\text{cos}^2$-pulse envelope with 30 cycles duration (10.9 cycles FWHM in intensity) and effective potentials taken from Refs. \cite{tong05a} and \cite{mueller1998} in the case of neon and argon, respectively. The latter is known to provide the most accurate results for argon, although the Cooper minimum predicted by this potential lies a few electron-Volts below its experimentally established position \cite{woerner09a,higuet11a}. The TDSE is propagated using the pseudo-spectral method described in \cite{tong1997,murakami2013}. The outermost sub-shell of the ground state consists of three degenerate $p$ orbitals: $p_+$, $p_-$ and $p_0$, where the index indicates the z-component of the orbital angular momentum. Since the $p_0$ orbital has a node in the polarization plane, its contribution to the harmonic signal is negligible. Therefore the contributions from the $p_+$ and $p_-$ orbitals are summed coherently to obtain harmonic spectra and ratios between neighboring harmonics. For each orbital, the harmonic signal is obtained from the Fourier-transformed time-dependent dipole acceleration.

We develop a deeper understanding of the observed effects by turning to calculations based on the SFA \cite{lewenstein94a}, which was extended to bi-circular driving fields in Ref. \cite{milosevic00a}. In order to capture system-specific effects, the photorecombination matrix elements traditionally evaluated in the plane-wave approximation in the SFA, are replaced by numerically exact matrix elements based on outgoing scattering waves $\Psi^{(+)}$ of the same model potentials as used in the TDSE \cite{morishita2008,woerner09a}. The application of the saddle-point approximation \cite{lewenstein94a,milosevic00a} leads to a three-step model consisting of a sequence of ionization, propagation and recombination \cite{corkum1993,frolov2009}. As a consequence of the combined spatio-temporal symmetries of the system and the laser field the total induced dipole moment contains contributions of three equivalent electron trajectories per optical cycle of the fundamental laser field \cite{mauger2016}. Harmonics of orders $3q+1$ are polarized as the fundamental and harmonics of orders $3q+2$ are polarized as the second harmonic \cite{becker1999,milosevic00a}. Therefore the dipole operator $\textbf{d}$ is projected on the relevant polarization vector $\textbf{e}_{\pm}=\textbf{e}_x\pm\I\textbf{e}_y$. We sum coherently over the contributions of the two initial atomic orbitals $p_+$ and $p_-$ with wave functions denoted as $\psi_{m=1}$ and $\psi_{m=-1}$ in what follows. Due to phase-matching conditions for the interaction region behind the laser focus, only the shortest trajectory has a significant contribution to the HHG signal. We have confirmed this statement by a numerical analysis of the macroscopic signal where we apply the theoretical methods described in Refs. \cite{salieres96a,balcou97a,gaarde08a} to bi-circular fields.
This leads to the following expression for the harmonic intensity:
\begin{eqnarray}\label{model}
I_{3q\pm1}= &9& \, \left|P(\boldsymbol{k}_s,t_s,t'_s)\right|^2 \\ \nonumber
 &\times& \left|\sum_{m=\pm1} d_{\textrm{rec},m}^{\pm}(\textbf{v}(\textbf{k}_s,t_s)) d_{\textrm{ion},m}(\textbf{v}(\textbf{k}_s,t'_s))\right|^2, 
\end{eqnarray}
with the complex-valued time of ionization $t_s'$, recombination $t_s$ and momentum ${\bf k}_s=-\int_{t'_s}^{t_s} \mathrm{d}t''\textbf{A}(t'')/(t_s-t'_s)$. The velocity of the electron is $\textbf{v}(\textbf{k},t)=\textbf{k}+\textbf{A}(t)$ with $\textbf{A}(t)=-\int^t\textbf{E}(t')\mathrm{d}t'$. Here and in what follows, we use atomic units unless otherwise stated. The times of ionization $t'_s$ and recombination $t_s$ are given as solutions of the corresponding saddle-point equations
\begin{align}
\textbf{v}(\textbf{k}_s,t'_s)^2/2\,=\,&-I_p \label{condion} \\ 
\textbf{v}(\textbf{k}_s,t_s)^2/2\,=\,& n\omega-I_p, \label{condrec}
\end{align} 
with the harmonic order $n$ and the ionization potential $I_p$.
All factors that are independent of the atomic orbitals are collected in the prefactor $P(\boldsymbol{k}_s,t_s,t'_s)$. The first transition matrix element $d_{\textrm{rec},m}^{\pm}=\langle \psi_m | \textbf{e}_{\pm}^{*}\cdot \textbf{d}| \Psi_{\textbf{v}(\textbf{k}_s,t_s)}^{(+)}\rangle$ describes the recombination step and the second one $d_{\textrm{ion},m}=\langle \textbf{v}(\textbf{k}_s,t'_s)|\psi_m\rangle$ describes the ionization step. The continuum states in the ionization process (only) are described as plane waves $|\textbf{v}\rangle$ \cite{perelomov1966,barth2011}. Electron tunnelling entirely takes place in imaginary time and results in a complex velocity $\textbf{v}(\textbf{k}_s,t'_s)$ of the electron at $t'_s$. Electron propagation in the laser field after ionization takes place in real time and results in the complex-valued velocity vector $\textbf{v}(\textbf{k}_s,t_s)$ at $t_s$. The imaginary part of the recombination velocity does not vanish in the adiabatic limit ($I_p\rightarrow 0$). Over the plateau of the harmonic spectrum the prefactor in Eq. (\ref{model}) is only weakly energy dependent. Hence it approximately cancels out in the calculation of intensity ratios between neighboring harmonic orders and was neglected to obtain the dotted curves in Fig. 2. 

The agreement between the TDSE (full red line in Fig. 2a) and the experiment is very good in the case of neon. We note that the intensity ratios between neighboring harmonics strongly depend on the intensity ratio $I(\omega)/I(2\omega)$ of the fundamental and its second harmonic. Remaining discrepancies between experiment and theory may thus be partially attributed to the limited accuracy to which $I(\omega)/I(2\omega)$ in the generation region is known. In the case of argon, the agreement between the experiment and the TDSE results is also good, especially in the high-energy region, after a global shift of $+6.2$~eV has been applied to the theoretical results. This global shift is consistent with earlier work \cite{woerner09a,le09a,morishita08a} and can therefore be attributed to limitations of the effective-potential approach in describing high-harmonic spectra.
 
The results of the SFA calculations agree well with those of the TDSE calculations and the experimental data in the high-energy part of both spectra, but the SFA systematically over-estimates the intensity ratios at low photon energies. A detailed analysis suggests that this discrepancy mainly originates from an overestimation of the asymmetry of the recombination step, that we attribute to a failure of the saddle-point approximation at very low kinetic energies.

The contributions of each of the ionization and recombination matrix elements are separately shown in Fig. 3. The panels (a) show the magnitudes of the ionization matrix elements, evaluated for the complex velocities at ionization as a function of the emitted photon energy. Interestingly, we find that the orbitals representing electrons co-rotating with the fundamental laser field dominate for low emitted photon energies, whereas the counter-rotating electrons dominate the contribution at high photon energies. Similar results have been obtained in \cite{ayuso16a}. The exact crossing point of the two curves depends on the intensity ratio of the two components of the bi-circular field and the ionization potential of the atom. It shifts to higher photon energies with increasing relative intensity of the second-harmonic field. 

\begin{figure}[t]
\includegraphics[width=0.5\textwidth]{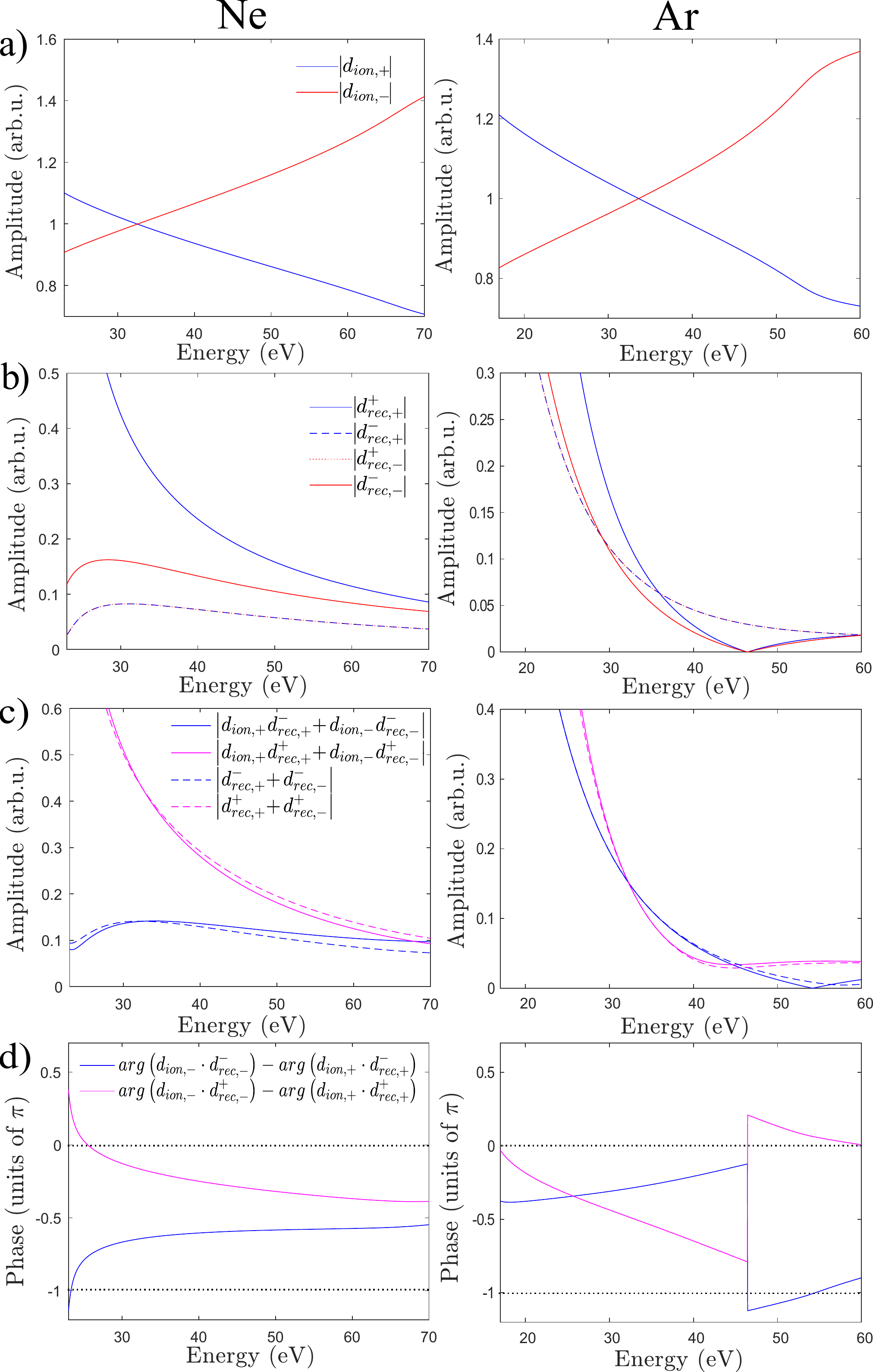}
\caption{(color online) a) Amplitudes of the matrix elements for ionization of $p_+$ and $p_-$ orbitals by bi-circular laser fields. The wavelengths and intensities are the same as those used in Fig. 2 and are given in the caption of Fig. 1, b) Amplitudes of the photorecombination matrix elements, c) Amplitudes of the total induced dipole moments responsible for the emission of the symmetry-allowed harmonics 3$q$+1 (magenta) and 3$q$+2 (blue), d) relative phases of the contributions from $p_+$ and $p_-$ orbitals to the results of panels (c).}
\end{figure}

The panels (b) show the magnitudes of the recombination matrix elements. In the case of neon (left column), the matrix element describing recombination to the orbital co-rotating with the fundamental ($p_+$) under emission of a photon of co-rotating polarization (${\bf e}_+$) dominates ($d^+_{rec,+}$), followed by recombination to the counter-rotating orbital under emission of a photon of counter-rotating polarization ($d^-_{rec,-}$). In the case of real-valued recombination velocities, these matrix elements would have equal moduli. The imaginary part of the recombination velocities is responsible for the observed differences between $d^+_{rec,+}$ and $d^-_{rec,-}$, which are very large at low energies in neon. We note that this effect is also responsible for the deviation of the ratios from unity in the case of the 1s-shell of helium, that were observed in \cite{baykusheva16a}. The ratio $\left|d^+_{rec,+}/d^-_{rec,-}\right|$ is found to decrease with increasing relative intensity of the second-harmonic field, for both Ne and Ar. The smaller amplitudes of the $d^+_{rec,-}$ and $d^-_{rec,+}$ matrix elements are expected because of the Fano-Bethe propensity rules \cite{fano85a,medisauskas15a}. 

In the case of argon (Fig. 3b, right), the dominance of $d^+_{rec,+}$ and $d^-_{rec,-}$ over $d^+_{rec,-}$ and $d^-_{rec,+}$ only holds below 30 eV. The breakdown of the Fano-Bethe propensity rule is caused by the fact that the $d^+_{rec,+}$ and $d^-_{rec,-}$ matrix elements go through zero at 46.4~eV. Since these two matrix elements contain only the contribution of the $\epsilon d\rightarrow 3p$ transition, the zero crossing is a direct consequence of the sign reversal of the $\epsilon d\rightarrow 3p$ radial matrix element at this energy. The $d^+_{rec,-}$ and $d^-_{rec,+}$ matrix elements, in contrast, additionally contain a contribution from the $\epsilon s \rightarrow 3p$ transition matrix element, which varies monotonically with energy and does not change sign. We note that the argon results in Fig. 3 have not been shifted in energy, in contrast to the theoretical results shown in Fig. 2b (see caption).

Figure 3c shows the magnitude of the total induced dipole moment for the harmonics of orders $3q+1$ (${\bf e}_+$-polarized, magenta) and $3q+2$ (${\bf e}_-$-polarized, blue) as full lines. In addition to the amplitude effects shown in panels (a) and (b), these results are influenced by the relative phase of the emission from the $p_+$ and $p_-$ orbitals, which is shown in the panels (d). The strong emission of $3q+1$ harmonics at low photon energies in neon originates from the large magnitude of $d^+_{rec,+}$ on one hand and the constructive interference between emissions from the $p_+$ and $p_-$ orbitals on the other hand, as revealed by panel (d). Similarly, the much weaker emission of the $3q+2$ harmonics is caused by the small magnitude of $d^-_{rec,-}$ and the destructive interference at low energies. These effects result in the observed very large values of the intensity ratio at low energies in neon. 
We further investigate the relative importance of the ionization and recombination matrix elements by setting the ionization matrix elements to unity. This leads to the dashed curves shown in Fig. 3c. The effect of the unequal ionization amplitudes leads to a more rapid merging of the two full curves at high energies in neon.

The much smaller ratios in the case of argon, as compared to neon, and the two-fold inversion of the ratio as a function of the photon energy also become immediately apparent from Fig. 3c. Both inversions are largely unchanged by the effect of unequal ionization amplitudes. The higher-lying crossing of the dashed curves occurs exactly at the position where $d^+_{rec,+}=d^-_{rec,-}=0$. The full lines cross in the immediate vicinity of this point, which shows that the higher-lying inversion of the intensity ratio in argon lies very close to the position where the $\epsilon d\rightarrow 3p$ radial matrix element changes sign. 

We note that all qualitative observations made in the case of Ar, i.e. the two-fold inversion of the ratio and the mentioned dependencies of ionization and recombination amplitudes on the ratio of the driving fields, have also been made in the case of Xe (not shown). Our work thus suggests that the qualitative differences between Ne and the heavier rare gases originate in the existence of radial nodes in the $np$ atomic wave functions for $n\ge 3$, which are also the prerequisite for the existence of Cooper minima \cite{cooper62a}.

This study marks the beginning of a quantitative understanding of bi-circular high-harmonic spectra. 
We have shown that the intensity ratios of neighboring harmonic orders are an important observable in BHHS that is particularly sensitive to the electronic structure of the target. These ratios were found to be subject to propensity rules that reflect the angular momentum of the probed orbital. In particular, strong-field ionization in BHHS was found to favor orbitals describing electrons that rotate in the same direction as the fundamental field ($p_+$) for quantum orbits emitting low photon energies and the p$_-$ orbitals at high photon energies (Fig. 3a). This result appears to be general, at least for rare gas atoms. Photorecombination in BHHS was found to preferentially lead to the emission of circularly-polarized radiation of the same direction of rotation as the orbital (Fig. 3b). Our study has additionally highlighted the importance of interference between high-harmonic emission from $p_+$ and $p_-$ orbitals as illustrated in Fig. 3d. Finally, the ratios also strongly depend on the radial structure of the orbital wave function through the atom-specific photorecombination matrix elements.
The results summarized in this letter outline the foundation for the quantitative interpretation of BHHS, which can now be extended to molecules \cite{baykusheva16a}, to time-dependent electronic wave packets in neutral molecules \cite{kraus13b,baykusheva14a,walt17a} and to probing sub-cycle electronic dynamics, such as charge migration \cite{kraus15b}.

\begin{acknowledgments}
D.B. and H.J.W. acknowledge support from an ERC starting grant (project no. 307270-ATTOSCOPE) and the Swiss National Science Foundation (SNSF) through project no. 200021\_159875. S.B. and M.L. acknowledge support from the Deutsche Forschungsgemeinschaft in the frame of the Priority Programme Quantum Dynamics in Tailored Intense Fields. S.B. thanks the Studienstiftung des deutschen Volkes for financial support.
\end{acknowledgments}


\end{document}